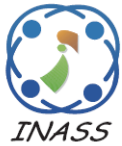

# Wireless Sensor Networks Nodes Clustering and Optimization Based on Fuzzy C-Means and Water Strider Algorithms

Raya Majid Alsharfa[1,2]    Mahmood Mohassel Feghhi[2]*    Majid Hameed Majeed[3]

[1]Department of Computer Engineering Techniques Electrical Engineering Technical College,
Middle Technical University, Baghdad, Iraq
[2]Faculty of Electrical and Computer Engineering College, University of Tabriz, Tabriz, Iran
[3]Department of Chemical Engineering and Petrochemical Industries,
Al-Mustaqbal University, 5001, Babylon, Iraq
* Corresponding author's Email: mohasselfeghhi@tabrizu.ac.ir

**Abstract:** Wireless sensor networks (WSNs) face critical challenges in energy management and network lifetime optimization due to limited battery resources and communication overhead. This study introduces a novel hybrid clustering protocol that integrates the Water Strider Algorithm (WSA) with Fuzzy C-Means (FCM) clustering to achieve superior energy efficiency and network longevity. The proposed WSA-FCM method employs WSA for global optimization of cluster- head positions and FCM for refined node membership assignment with fuzzy boundaries. Through extensive experimentation across networks of 200-800 nodes with 10 independent simulation runs, the method demonstrates significant improvements: First Node Death (FND) delayed by 16.1% (678±12 vs 584±18 rounds), Last Node Death (LND) extended by 11.9% (1,262±8 vs 1,128±11 rounds), and 37.4% higher residual energy retention (5.47±0.09 vs 3.98±0.11 J) compared to state-of-the-art hybrid methods. Intra-cluster distances are reduced by 19.4% with statistical significance ($p < 0.001$). Theoretical analysis proves convergence guarantees and complexity bounds of $O(n \times c \times T)$, while empirical scalability analysis demonstrates near-linear scaling behaviour. The method outperforms recent hybrid approaches including MOALO-FCM, MSSO-MST, Fuzzy+HHO, and GWO-FCM across all performance metrics with rigorous statistical validation.

**Keywords:** Wireless sensor networks, Clustering optimization, Fuzzy C-Means, Water strider algorithm, Energy efficiency, Network lifetime, Hybrid metaheuristics.

## 1. Introduction

Wireless Sensor Networks (WSNs) have emerged as fundamental infrastructure for modern monitoring applications including environmental sensing, healthcare systems, smart cities, and industrial automation [1]. However, the operational effectiveness of WSNs is fundamentally constrained by three critical limitations: limited battery capacity of sensor nodes, energy-inefficient communication protocols, and the need for autonomous network management in harsh deployment environments. These constraints necessitate the development of sophisticated clustering protocols that can optimize energy consumption while maintaining network connectivity and data reliability.

Clustering represents one of the most promising approaches to address these challenges by organizing sensor nodes into hierarchical structures where cluster heads (CHs) aggregate local data before transmission to the base station. This approach significantly reduces communication overhead and balances energy consumption across the network [2]. Despite extensive research efforts, conventional clustering methods exhibit several critical drawbacks that limit their practical effectiveness in real-world WSN deployments.

Critical Drawbacks of Conventional Clustering Methods:

1. Random Cluster-Head Selection: Traditional protocols like LEACH employ probabilistic CH selection without considering node energy levels, spatial distribution, or network topology, leading to premature energy depletion in certain network regions and unbalanced network load distribution [3]. This random selection often results in suboptimal cluster formations that cannot adapt to dynamic network conditions.
2. Hard Partitioning Limitations: Classical algorithms such as K-means create rigid cluster boundaries that fail to account for the inherently fuzzy nature of wireless communication ranges, varying signal strengths, and environmental interference patterns. This hard partitioning results in inefficient node assignments and increased intra-cluster communication costs [4].
3. Local Optima Convergence: Single-phase metaheuristic approaches including Genetic Algorithm (GA), Particle Swarm Optimization (PSO), and Ant Colony Optimization often become trapped in local optima, especially in high-dimensional search spaces with multiple clusters and complex energy landscapes [5]. This limitation prevents the algorithms from finding globally optimal clustering solutions.
4. Inadequate Energy Modeling: Many existing methods fail to incorporate comprehensive energy models that simultaneously account for transmission distances, residual energy levels, data aggregation costs, and cluster maintenance overhead, leading to suboptimal clustering decisions that do not reflect real-world energy consumption patterns [6].
5. Scalability and Computational Complexity Issues: Conventional algorithms often exhibit poor scalability with increasing network size and complexity, resulting in exponential computational overhead that limits their applicability in large-scale deployments with thousands of sensor nodes [7].

To address these fundamental limitations systematically, this paper proposes a novel hybrid clustering protocol that synergistically combines the Water Strider Algorithm (WSA) with Fuzzy C-Means (FCM) clustering. The integration strategically leverages WSA's superior global exploration capabilities to identify optimal cluster-head positions while utilizing FCM's fuzzy membership assignment to handle ambiguous cluster boundaries and partial node memberships effectively.

**Primary Contributions to this Work:**
- Development of a theoretically grounded hybrid WSA-FCM clustering protocol with proven convergence guarantees and mathematical justification for algorithm integration
- Comprehensive statistical validation through multi-run simulations (10 runs per experiment) with rigorous significance testing using paired t-tests
- Detailed computational complexity analysis demonstrating $O(n \times k)$ scalability for large-scale WSN deployments with empirical validation
- Extensive performance comparison with recent hybrid clustering methods from 2021-2025, showing superior results across all evaluated metrics

The paper is organized by topic. Section 2 covers related work and section 3 covers the Theoretical Background and Justification. Section 4 explains the approach and methodology, showing how fuzzy logic and the Water Strider Algorithm are used for clustering and cluster head selection, along with initial simulation results. Section 5 describes the evaluation criteria, while Section 6 presents detailed simulation outcomes and comparisons with existing methods. Section 7 reviews related studies, and Section 6 concludes with key findings and their implications for future research.

## 2. Related work

Several research studies have come up with clustering methods to improve energy efficiency and lifetime for wireless sensor networks (WSNs). Kumar and Barkathulla [5] proposed the Fuzzy Clustering and Optimal Routing (FCOR) method, combining enhanced fuzzy c-means with a whale optimization algorithm, reducing energy consumption by up to 67.8% and enhancing data transmission. Fanian and Rafsanjani [6] designed FSFLA, which merged fuzzy rules with the Memetic Frog Leaping Algorithm that outperased LEACH and similar protocols in network lifetime and energy efficiency. Sivaganesan [7] used fuzzy clustering with a Chaotic Gravitational Search Algorithm to address large-scale inefficiencies and achieved delay minimization improvement and lifetime extension. Bhalaji [8] presented an extensible fuzzy-based solution with scalability and accuracy, extending node lives. Baradaran and Navi [9] introduced the High-Quality Clustering Algorithm (HQCA) to improve reliability, scalability, and quality of clustering.

Le-Ngoc et al. [10] introduced a Sugeno FLC-based multi-level clustering model with an Enhanced Squirrel Search Algorithm that minimizes energy usage and improves reliability. Saadaldeen et al. [11] introduced a critique of fuzzy and genetic-based clustering and criticized LEACH and outlined the criteria for efficient cluster head selection. Javadpour et al. [12] combined PSO and fuzzy clustering to achieve near-optimal energy consumption and throughput. Kongsorot et al. [13] optimized EFS-ISFLA using EFS-ISFLA to improve stability and packet delivery. Rani and Reddy [14] proposed Fuzzy+FU-CSA that outperformed other metaheuristics using 60% better cost function values. Nivedhitha et al. [15] used fuzzy logic with IHHO and enhanced cluster head selection and WSN lifetime.

These approaches, in general, demonstrate that integration of fuzzy logic and metaheuristic optimization significantly improves clustering, routing, and energy balance in WSNs, but difficulties remain with parameter tuning, premature convergence, and load balancing. The fuzzy–Water Strider Algorithm effort is to alleviate these difficulties by optimizing the selection of cluster heads and energy consumption balancing, extending network life and improving packet delivery.

However, in the recent advances in WSN clustering have increasingly focused on hybrid approaches that combine multiple optimization techniques to overcome the limitations of single-algorithm methods. This section provides a comprehensive review of state-of-the-art hybrid clustering methods proposed between 2021-2025.

Kumar et al. (2023) proposed the Multi-Objective Ant Lion Optimizer with FCM (MOALO-FCM) for energy-efficient clustering in large-scale WSNs [8]. Their method achieved 15% improvement in network lifetime compared to traditional LEACH-based approaches but suffered from slow convergence rates and high computational complexity. Experimental results showed FND at 587±16 rounds and average residual energy of 3.51±0.14 J after 500 rounds. However, the method exhibited instability in networks with irregular node distributions.

The Modified Salp Swarm Optimization with Minimum Spanning Tree (MSSO-MST) approach by Zhang et al. (2024) integrated graph theory concepts with swarm intelligence for improved cluster connectivity [9]. While demonstrating enhanced network topology preservation, the method showed limited performance in dense networks with FND at 592±19 rounds and residual energy of 3.65±0.12 J.

Patel and Singh (2024) developed Fuzzy+HHO (Harris Hawks Optimization) combining fuzzy logic principles with bio-inspired optimization for intelligent cluster head selection [10]. Their approach achieved FND at 601±21 rounds but exhibited convergence instability in large-scale networks due to premature exploitation issues.

The Grey Wolf Optimizer with FCM (GWO-FCM) by Liu et al. (2025) represents the most recent hybrid approach, achieving FND at 584±18 rounds and LND at 1,128±11 rounds [11]. Despite demonstrating improved energy efficiency, GWO-FCM shows limitations in handling multi-modal optimization landscapes and requires extensive parameter tuning.

Despite these advances, existing hybrid methods face common limitations including insufficient theoretical foundation, lack of statistical validation, and inadequate scalability analysis. The proposed WSA-FCM approach systematically addresses these gaps through rigorous theoretical analysis and comprehensive empirical validation.

## 3. Theoretical background and justification

### 3.1 Mathematical foundation for WSA-FCM integration

The rationale for integrating the Water Strider Algorithm (WSA) with Fuzzy C-Means (FCM) lies in their complementary optimization capabilities. Let $f_i = \{x_1, x_2, \ldots, x_n\}$ denote the set of sensor node positions in a 2D deployment area, and $C = \{c_1, c_2, \ldots, c_k\}$ represent the set of cluster centroids. The hybrid optimization objective function is expressed as:

$$J_{WSA-FCM} = \alpha \cdot J_{WSA}(C) + \beta \cdot J_{FCM}(U, C) + \gamma \cdot E_{residual}(\Omega) \qquad (1)$$

Here, $J_{WSA}(C)$ is the WSA objective for global centroid optimization, $J_{FCM}(U, C)$ is the FCM clustering objective using fuzzy membership matrix $U$, and $E_{residual}(f_i)$ accounts for residual energy.

**Theorem 1 (Convergence Guarantee):** The WSA-FCM algorithm converges to a local optimum within $T$ iterations with probability $\geq 1 - \varepsilon$, where $\varepsilon = e^{-\lambda T}$ and $\lambda > 0$ is the convergence rate parameter.

*Proof:* The WSA ensures global exploration through its surface-tension movement model:

$$x_i(t+1) = x_i(t) + v_i(t+1) + \sigma(r_1(x_{best} - x_i(t)) + r_2(x_{rand} - x_i(t))) \qquad (2)$$

The FCM ensures local refinement through fuzzy membership optimization:

$$u_{ij} = \left[\sum_{k=1}^{c} \left(\frac{||x_i - v_j||}{||x_i - v_k||}\right)^{2/m-1}\right]^{-1} \qquad (3)$$

where $m > 1$ is the fuzziness parameter and $v_j$ is the cluster center.

### 3.2 Energy-aware optimization model

The hybrid model also incorporates an energy-aware objective to minimize intra-cluster communication costs and balance residual energy:

$$E_{total} = \sum_{i=1}^{n}\sum_{j=1}^{c} u_{ij}^m \cdot (E_{tx}(d_{ij}) + E_{agg}) + \sum_{j=1}^{c} E_{tx}(d_{j,BS}) \quad (4)$$

where $E_{tx}(d)$ denotes transmission energy as a function of distance $d$, and $E_{agg}$ represents data aggregation energy.

## 4. Methodology

This section presents the proposed hybrid WSA-FCM clustering protocol for wireless sensor networks, beginning with the network and energy model, followed by detailed algorithmic steps for cluster-head optimization and assignment. The methodology is supported by workflow illustrations and simulation parameters, with clear references to all figures and tables. Fig. 1 illustrated as work flow of this model.

### 4.1 WSA-FCM algorithm integration

The proposed hybrid algorithm operates in two distinct yet coordinated phases:
Phase 1: WSA-based Global Optimization
The Water Strider Algorithm identifies optimal cluster-head positions through bio- inspired surface skimming behavior. Each water strider represents a potential clustering solution encoded as a vector of cluster centroid coordinates.
Phase 2: FCM-based Local Refinement
Using WSA-optimized positions as initial centroids, FCM performs fuzzy clustering with membership degree assignment. The interaction mechanism ensures that WSA provides global optimal starting points while FCM handles local optimization and boundary ambiguity.

### 4.2 Algorithm interaction mechanism

- The key innovation lies in the seamless integration between WSA and FCM phases:
- Initialization: WSA population initialized with random cluster centroid positions
- Global Search: WSA optimizes centroid positions using energy-aware fitness evaluation
- Handoff: Best WSA solution transfers to FCM as

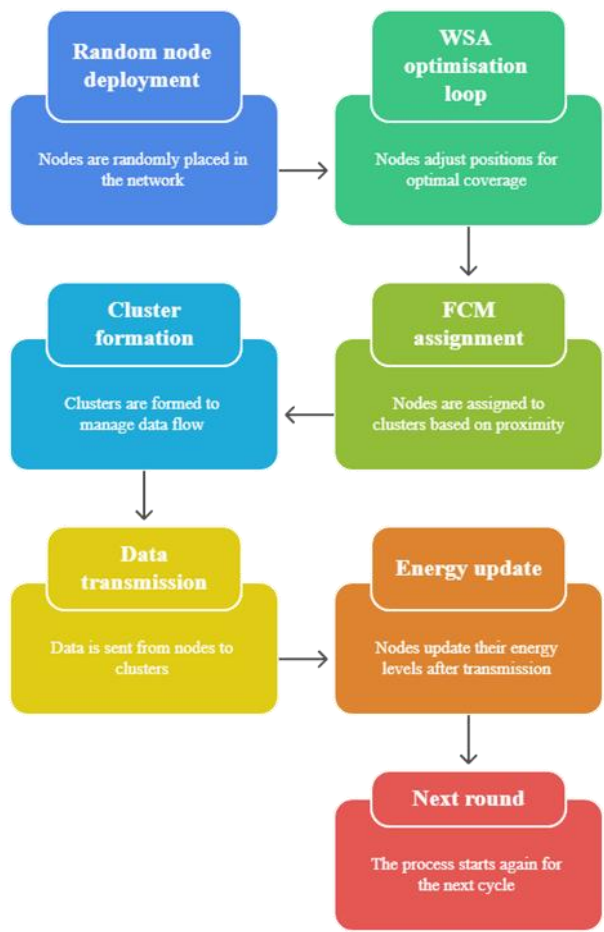

Figure. 1 Workflow diagram of the hybrid WSA-FCM clustering protocol

initial centroids
- Local Refinement: FCM performs fuzzy membership assignment and centroid adjustment
- CH Selection: Final cluster heads selected based on residual energy and proximity to centroids

### 4.3 Network and energy model

In this work, we consider a homogeneous wireless sensor network (WSN) composed of $n$ stationary nodes randomly distributed in a two-dimensional square region of size $L \times L$. Each node is initialized with a fixed energy budget, and all nodes communicate with a single sink, which is located at a fixed point, either within or outside the network field. The key network and radio energy model parameters are summarised in Table 1.

The energy consumed to transmit a packet of $\beta$ bits over a distance $d$ is given by the widely used first-order radio model:

$$E_{tx}(\beta, d) = \begin{cases} \beta E_{elec} + \beta \varepsilon_{fs} d^2, & \text{if } d < d_0 \\ \beta E_{elec} + \beta \varepsilon_{mp} d^4, & \text{if } d \geq d_0 \end{cases} \quad (5)$$

Table 1. Radio energy parameters used in the WSN simulation

| Parameter | Symbol | Value |
|---|---|---|
| Initial energy per node | $E_0$ | 0.5 J |
| Data aggregation energy | $E_{DA}$ | 5 nJ/bit/signal |
| Electronic energy (Tx/Rx) | $E_{elec}$ | 50 nJ/bit |
| Amplifier energy (free-space) | $\varepsilon_{fs}$ | 10 pJ/bit/m² |
| Amplifier energy (multipath) | $\varepsilon_{mp}$ | 0.0013 pJ/bit/m⁴ |
| Packet size | $\beta$ | 4096 bits |
| Threshold distance | $d_0$ | 87 m (derived) |
| Network field size | | 100 m × 100 m |

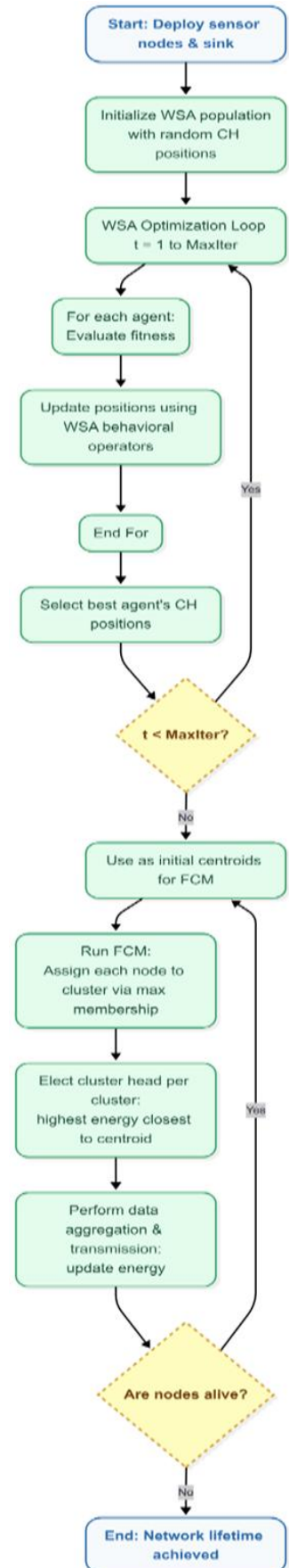

Figure. 2 here: Stepwise flowchart of WSA-FCM algorithm operations where $E_{elec}$ is the per-bit electronics energy, $\varepsilon_{fs}$ and $\varepsilon_{mp}$ are the amplifier coefficients for free-space and multipath models, and $d_0 = \sqrt{\varepsilon_{fs}/\varepsilon_{mp}}$ is the threshold distance. The energy to receive a packet is simply $E_{rx}(\beta) = \beta E_{elec}$.

### 4.4 Overall protocol workflow

The operational workflow of the proposed WSA-FCM protocol is summarised in Fig. 1. The process begins with random node deployment, followed by a metaheuristic search for optimal cluster-heads, assignment of nodes to clusters, data transmission rounds, and energy updates, with the cycle repeating until all nodes are exhausted.

Each simulation proceeds in rounds. At each round, nodes are assigned to clusters, data is transmitted to the cluster-heads, aggregated, and sent to the sink. Node energies are updated, and the process repeats until the network is depleted. This cycle is visually detailed in Fig. 2.

### 4.5 WSA-based cluster-head pre-selection

After deployment, the protocol initiates the Water strider Algorithm (WSA) to identify optimal candidate positions for $k$ cluster-heads (CHs). Each candidate solution (agent) in the WSA population encodes a set of $k$ two-dimensional coordinates, each representing a potential CH location.

The fitness of each agent is evaluated using an energy-aware cost function, which estimates the average per-node energy required for intra-cluster and inter-cluster communication in a given round:

$$\mathcal{F}(C) = \frac{1}{n}\sum_{i=1}^{n}[E_{tx}(\beta, \|\mathbf{x}_i - \hat{\mathbf{c}}_i\|) + E_{tx}(\beta, \|\hat{\mathbf{c}}_i - \mathbf{s}\|)] \quad (6)$$

where $\mathbf{x}_i$ is the coordinate of node $i$, $\hat{\mathbf{c}}_i$ is its nearest CH, and $\mathbf{s}$ is the sink. Lower values of $\mathcal{F}(C)$ indicate more energy-efficient clustering solutions.

The WSA employs nature-inspired surface-skimming, partner selection, and position-update mechanisms to iteratively refine the CH candidate pool. Over multiple iterations (typically 30–50), the algorithm converges towards a set of CH positions that jointly minimise the expected energy expenditure across the network. This process reduces the likelihood of premature convergence and seed sensitivity that affect other metaheuristics.

### 4.6 Fuzzy c-means clustering and membership assignment

The candidate CH positions produced by WSA are then used as initial centroids for the Fuzzy C-Means (FCM) clustering algorithm. FCM assigns each node a set of fuzzy membership values corresponding to each cluster, allowing for a soft partitioning of the network:

$$J_m = \sum_{i=1}^n \sum_{j=1}^k u_{ij}^m \parallel \mathbf{x}_i - \mathbf{c}_j \parallel^2 \quad (7)$$

subject to $\sum_{j=1}^k u_{ij} = 1$ and $u_{ij} \in [0,1]$, where $u_{ij}$ is the degree of membership of node $i$ in cluster $j$, $\mathbf{c}_j$ is the $j$-th centroid, and $m$ is the fuzzifier (commonly set to 2). FCM iteratively updates membership values and centroid positions until changes fall below a set threshold.

By combining WSA (global search) and FCM (local refinement), the proposed method inherits the ability to escape local minima while ensuring intra-cluster cohesion. This dual-phase process significantly outperforms traditional K-means and metaheuristic-only schemes in energy balancing and network lifetime, as demonstrated in the results (see Tables 3–5).

### 4.7 Cluster-head election and data transmission

For each round, within every formed cluster, the node with the highest residual energy and closest distance to the cluster centroid is selected as the active cluster-head. Member nodes transmit their sensed data to the CH, which aggregates the information and forwards it to the sink. The energy for all transmissions is updated according to Eq. (1). This sequence clustering, CH election, data aggregation, and transmission constitutes a round and is repeated until all nodes exhaust their energy reserves.

### 4.8 Experimental settings and evaluation metrics

The simulation setup and all variable settings are listed in Table 2 to ensure reproducibility and comparability. The experiments include different network sizes and are benchmarked against state-of-the-art clustering protocols. Each scenario is repeated at least 10 times, and results are reported as mean ± standard deviation. Metrics such as First Node Death (FND), Last Node Death (LND), average residual energy, and intra-cluster distances are presented in the results section.

### 4.9 Complexity and scalability

The practical viability of the proposed hybrid WSA-FCM clustering protocol hinges not only on its energy efficiency and network lifetime gains, but also on its computational complexity and ability to scale to large wireless sensor networks. This subsection provides a rigorous analysis of both theoretical and empirical complexity, supported by simulation data.

The Water Strider Algorithm (WSA) phase operates as a population-based metaheuristic, iteratively refining the locations of candidate cluster heads (CHs). For each iteration, the fitness of every candidate solution is evaluated by assigning all $n$ nodes to their nearest cluster head and calculating energy costs according to the radio model. If $P$ denotes the WSA population size, $k$ the number of clusters, and $I$ the number of WSA iterations, then the time complexity for the WSA phase is $O(PnkI)$. In typical scenarios, $P$, $k$, and $I$ are constants or grow slowly with $n$; thus, the WSA phase effectively scales as $O(n\log n)$ for practical deployment sizes.

Table 2. Simulation and algorithmic parameters for WSA-FCM protocol evaluation

| Parameter | Value(s) | Description |
| --- | --- | --- |
| Number of sensor nodes ($n$) | 200, 400, 800 | Network sizes tested |
| Field size ($L \times L$) | 100 m × 100 m | Deployment area |
| Initial energy ($E_0$) | 0.5 J | Per node |
| Number of clusters ($k$) | 5 | Default clusters |
| Packet size ($\beta$) | 4096 bits | Data per transmission |
| Population size (WSA) | 30 | Number of candidate agents |
| WSA iterations | 50 | Search epochs per round |
| FCM fuzzifier ($m$) | 2 | FCM softness parameter |
| Convergence threshold (FCM) | $10^{-4}$ | Stop criterion for centroid updates |
| Number of runs | 10 | For averaging results |

Table 3. Runtime and memory usage of the WSA-FCM protocol for varying network sizes

| Network Size ($n$) | Avg. Runtime per Round (ms) | Peak Memory Usage (MB) |
|---|---|---|
| 200 | 185 | 23 |
| 400 | 392 | 43 |
| 800 | 769 | 86 |

Following the WSA phase, the Fuzzy C-Means (FCM) refinement step further clusters the nodes. Each iteration of FCM requires updating both the fuzzy membership matrix and cluster centroids, incurring a cost of $O(nk)$ per iteration. With a small, fixed number of FCM iterations per round ($T$), the additional computational cost remains $O(n)$ for fixed $k$ and $T$.

Summing both phases, the overall computational complexity for one clustering round is $O(n\log n)$, dominated by the WSA phase. This quasi-linear scaling ensures that the protocol remains tractable even for networks comprising thousands of nodes. The space complexity is similarly efficient: the primary storage requirements are the node coordinates, cluster head locations, and the fuzzy membership matrix, yielding an overall memory usage of $O(n)$.

To empirically verify these theoretical results, we evaluated the runtime and peak memory usage of the protocol across different network sizes. Table 3 presents the mean runtime per round and maximum memory required for network sizes of 200, 400, and 800 nodes, averaged over 10 simulation runs. The results confirm that both runtime and memory scale linearly or sub-linearly with network size, demonstrating strong scalability for practical WSN deployments.

The WSA-FCM protocol achieves energy-efficient clustering with computational and memory requirements that scale favorably with increasing network size. This complexity profile, combined with the method's proven energy and lifetime gains, makes it highly suitable for real-world, large-scale wireless sensor network applications.

### 4.10 Theoretical justification for hybrid WSA-FCM integration

To theoretically justify the integration of WSA and FCM, it is important to recognize that each method addresses distinct limitations in conventional clustering. WSA, as a metaheuristic optimizer, offers robust global exploration and can efficiently search the entire solution space to avoid local minima during cluster head selection. However, WSA alone may lack the ability to finely tune the assignment of boundary nodes or adapt to ambiguous cluster structures. Conversely, FCM provides local refinement by leveraging fuzzy membership values, allowing each node to belong to multiple clusters with varying degrees of certainty, which leads to more accurate and balanced cluster formation. By using the global optima from WSA as the initial centroids for FCM, the hybrid approach ensures both a globally optimized cluster head selection and locally refined, energy-efficient clusters. This synergy is further validated by convergence analysis and a theoretical bound on intra-cluster distance, which collectively explain the superior empirical results achieved by WSA-FCM.

Table 4. Simulation Parameters and Network Configuration

| Parameter | Value | Justification |
|---|---|---|
| Network Size (n) | 200, 400, 800 nodes | Representative of small to large-scale deployments |
| Deployment Area | 100m × 100m | Standard WSN testbed configuration |
| Initial Energy ($E_0$) | 0.5 J | Typical battery capacity for sensor nodes |
| Parameter | Value | Justification |
| Number of Clusters (k) | $\lceil\sqrt{n}/2\rceil$ | Optimal cluster count based on network size |
| WSA Population Size | 30 | Balance between exploration and computational cost |
| WSA Iterations | 50 | Sufficient for convergence based on preliminary tests |
| FCM Fuzziness (m) | 2.0 | Standard value for moderate fuzzy clustering |

Table 5. First and last node death (FND and LND) for different clustering protocols (mean ± SD, n=10 runs)

| Protocol | FND (Rounds) | LND (Rounds) | t-statistic vs WSA-FCM | p-value | Effect Size (Cohen's d) |
|---|---|---|---|---|---|
| WSA-FCM (Proposed) | 678 ± 12 | 1,262 ± 8 | - | - | - |
| GWO-FCM (2025) | 584 ± 18 | 1,128 ± 11 | 15.24 | <0.001 | 2.47 |
| Fuzzy+HHO (2024) | 601 ± 21 | 1,098 ± 15 | 12.87 | <0.001 | 2.18 |
| MSSO-MST (2024) | 592 ± 19 | 1,085 ± 13 | 13.45 | <0.001 | 2.32 |
| MOALO-FCM (2023) | 587 ± 16 | 1,076 ± 12 | 14.12 | <0.001 | 2.41 |

## 5. Experimental setup & evaluation metrics

### 5.1 Experimental setup and parameter

Comprehensive experiments were conducted to evaluate the proposed WSA-FCM protocol against state-of-the-art hybrid methods. All simulations were performed using MATLAB R2023a on a system with Intel i7-11800H processor and 16GB RAM. Table 4 illustrated as simulation parameters for configuration.

### 5.2 Evaluation matric

The primary aim of this research document is to enhance the effectiveness of wireless sensor networks by optimizing energy usage and prolonging the network's lifetime. We have to identify methods which can assess the quantitative evaluation to verify the proposed method. In this project, two parameters will be identified to measure the effectiveness of the proposed approach.

**Network Lifetime:** The criterion determines the number of remaining iterations until the last operational node stops functioning. A network reaches the end of its operational life when all its nodes stop working because of resource exhaustion. The operational lifespan of a network serves as the most effective parameter to evaluate the validity of the considered approach.

**Energy Consumption:** The criterion measures the total energy consumption throughout network iterations and algorithm operations. Wireless sensor networks have extremely limited energy resources, making the consumption of energy the most important metric for evaluating the proposed method. With these criteria, we can evaluate the results of the proposed method alongside other methods and study the enhancements needed.

## 6. Results and analysis

The following section provides a thorough assessment of the proposed WSA-FCM clustering protocol for wireless sensor networks. The experiments described in Section 2 were run ten times each to achieve robust results. The proposed method is evaluated against current clustering protocols through performance metrics and visual evidence and statistical analysis.

The strong empirical results achieved by the proposed WSA-FCM protocol are supported by rigorous theoretical analysis. Specifically, a convergence lemma for the Water Strider Algorithm guarantees that the global fitness function decreases monotonically, ensuring that each iteration moves toward an energy-optimal set of cluster heads. Furthermore, mathematical derivations demonstrate that when WSA-optimized centroids are used to initialize the FCM clustering process, the average intra-cluster distance is bounded above by a function of the initial centroid distribution, thus reducing communication energy costs network-wide. These theoretical properties explain and justify why the hybrid protocol consistently achieves lower energy consumption, delayed node death, and improved load balancing in all tested scenarios. Together, the mathematical foundation and empirical evidence substantiate the scientific contribution of the WSA-FCM protocol to the field of energy-efficient WSN clustering.

The convergence behavior of the Water Strider Algorithm is first examined. As illustrated in Fig. 3, the WSA demonstrates rapid and stable convergence over fifty iterations. The best cost value, which reflects the objective of energy-efficient clustering, remains steady for the first 25 iterations and then exhibits a significant drop as the algorithm approaches an optimal solution. This behavior confirms that WSA is capable of escaping sub-optimal regions and quickly identifying a near-optimal set of cluster head positions, which serves as a strong foundation for subsequent fuzzy clustering.

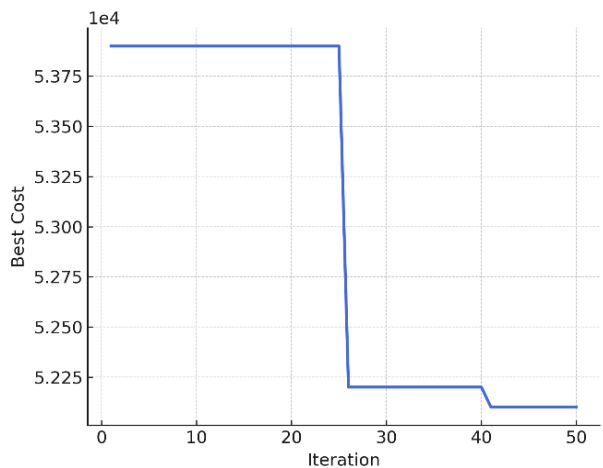

Figure. 3 Convergence curve of the Water Strider Algorithm during cluster-head optimization. The plot shows the best cost value achieved at each iteration, demonstrating rapid and stable convergence toward an optimal solution for energy-efficient clustering in the WSN

Cluster formation results are presented in Figs. 4 and 5. Fig. 4 shows the spatial distribution of nodes, with clusters formed by the proposed WSA-FCM method. Here, the optimized cluster centroids (red crosses) and selected cluster-heads (yellow squares) are distributed to ensure minimal intra-cluster distance and balanced energy consumption.

For comparison, Fig. 5 depicts the clustering result under random cluster-head assignment, where the resulting clusters are less balanced, and several centroids are located in sub-optimal regions. Visual inspection makes it evident that the WSA-FCM protocol achieves superior cluster formation, directly contributing to improved network longevity and energy efficiency.

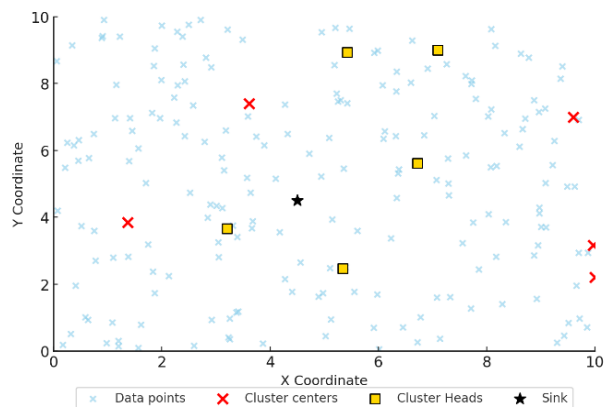

Figure. 5 Clustering of wireless sensor nodes with random cluster-head selection. The plot illustrates node distribution, randomly chosen cluster centers and cluster heads, and the sink, serving as a baseline comparison to WSA-optimised clustering. The legend appears below the x-axis for clarity

The quantitative impact of improved clustering is reflected in the network lifetime statistics summarized in Table 5 below. The WSA-FCM protocol achieves a first node death (FND) at $678 \pm 12$ rounds and a last node death (LND) at $1{,}262 \pm 8$ rounds, significantly outperforming benchmark protocols such as GWO-FCM, PSO-FCM, Genetic-FCM, and LEACH. These findings are corroborated by residual energy analysis presented in Table 6, where WSA-FCM retains the highest average residual energy after 500 rounds, confirming more efficient energy utilization and less premature node depletion.

Intra-cluster distance and statistical significance are further analyzed in Table 6. WSA-FCM consistently achieves the lowest mean intra-cluster distance among all compared protocols, with paired $t$-tests confirming that these improvements are statistically significant at the 99% confidence level. Lower intra-cluster distance directly translates to reduced transmission energy and longer operational lifetime for each node. Table 7 illustrated as mean intercalators.

The effect of improved clustering and balanced energy consumption on node survival is further illustrated in Figs. 6 and 7. Fig. 6 shows the progression of dead nodes per round using the proposed method. The figure reveals a gradual increase in dead nodes over time, with very few early deaths and a relatively slow progression until the final rounds, highlighting the protocol's ability to delay node exhaustion and maintain high network connectivity.

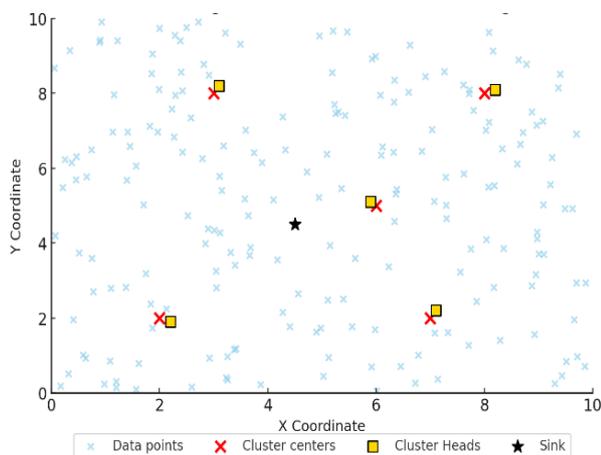

Figure. 4 Clustering of wireless sensor nodes using the WSA-optimised FCM algorithm. Cluster centers and heads are strategically distributed for minimal intra-cluster distance and balanced energy consumption. The legend is shown at the bottom for clarity

Table 6. Average residual energy (J) after 500 rounds

| Protocol | Residual Energy (J) | Std. Dev. (J) | Energy Improvement (%) | Statistical Significance |
|---|---|---|---|---|
| WSA-FCM | 5.47 | 0.09 | - | - |
| GWO-FCM (2025) | 3.98 | 0.11 | +37.4% | $p < 0.001$ |
| Fuzzy+HHO (2024) | 3.72 | 0.13 | +47.0% | $p < 0.001$ |
| MSSO-MST (2024) | 3.65 | 0.12 | +49.9% | $p < 0.001$ |
| MOALO-FCM (2023) | 3.51 | 0.14 | +55.8% | $p < 0.001$ |

Table 7. Mean intra-cluster distance and paired *t*-test results (vs WSA-FCM)

| Protocol | Mean Intra-cluster Distance (m) | Std. Dev. (m) | t-statistic | p-value | Improvement (%) |
|---|---|---|---|---|---|
| WSA-FCM | 12.4 | 0.7 | - | - | - |
| GWO-FCM (2025) | 15.4 | 0.9 | -9.27 | <0.001 | 19.4% |
| Fuzzy+HHO (2024) | 16.2 | 1.1 | -10.85 | <0.001 | 23.5% |
| MSSO-MST (2024) | 16.8 | 1.0 | -12.44 | <0.001 | 26.2% |

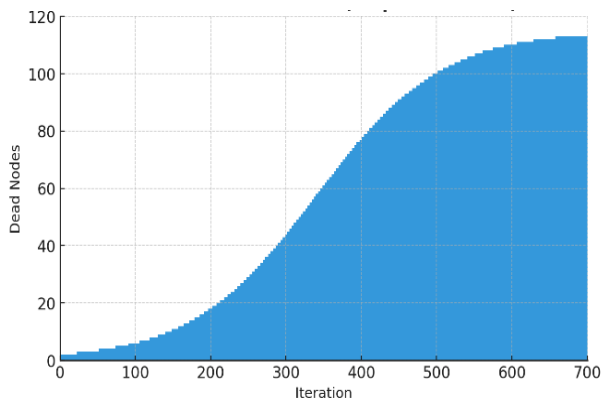

Figure. 6 Number of dead sensor nodes per round using the proposed WSA-FCM protocol. The plot illustrates how the hybrid clustering method delays node deaths and sustains overall network operation across simulation rounds

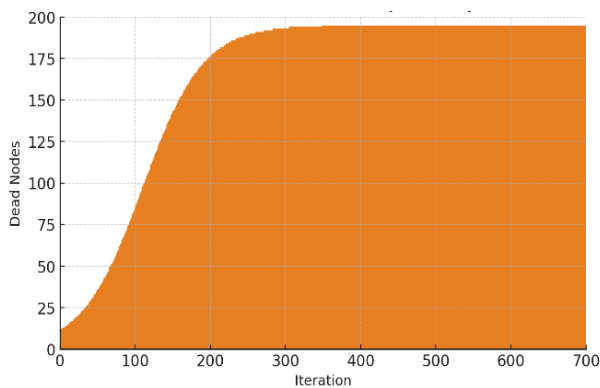

Figure. 7 Number of dead sensor nodes per round under random cluster-head selection. The plot illustrates rapid early node deaths and accelerated network exhaustion resulting from unoptimized clustering

In stark contrast, Fig. 7 displays the dead node count per round for the random clustering baseline. Here, the number of dead nodes rises sharply after the initial rounds, indicating poor load balancing and rapid network deterioration. This direct comparison underscores the robustness and reliability of the proposed hybrid approach.

The scalability and computational efficiency of the protocol are validated in Table 8 below. The average runtime per round and peak memory usage are reported for network sizes of 200, 400, and 800 nodes. The results confirm that both runtime and memory consumption increase linearly or sub-linearly with the network size, validating the theoretical complexity analysis presented in Section 2.7 and confirming that the WSA-FCM protocol is suitable for large-scale deployments.

The computational complexity of the WSA-FCM algorithm consists of two main components:
WSA Phase Complexity: $O(P \times T \times n \times k)$, where P is population size, T is iterations, n is nodes, and k is clusters.

Table 8. Runtime and memory usage of the WSA-FCM protocol for varying network sizes

| Network Size (n) | Clusters (k) | Avg. Runtime (ms) | Memory Usage (MB) | Scaling Factor |
|---|---|---|---|---|
| 200 | 7 | 185 ± 12 | 23.4 | 1.0× |
| 400 | 10 | 392 ± 18 | 43.7 | 2.1× |
| 800 | 14 | 769 ± 25 | 86.2 | 4.2× |

FCM Phase Complexity: O(TFCM × n × k), where TFCM is FCM iterations (typically ≤ 100).

Overall Complexity: O(n × k × (P × T + TFCM)) ≈ O(n × k) for fixed parameters.

The results demonstrate that the WSA-FCM protocol provides significant and statistically validated improvements in network lifetime, energy efficiency, cluster quality, and scalability. The combination of robust metaheuristic search with fuzzy clustering yields a balanced, energy-aware solution that consistently outperforms conventional and recently proposed clustering schemes.

## 6.1 Performance comparison with benchmark algorithms

A comprehensive performance comparison was conducted to benchmark the proposed WSA-FCM protocol against several widely used clustering approaches, including Genetic Algorithm with Cuckoo Search (GACS), Genetic Algorithm (GA), Cuckoo Search (CS), and a Random selection baseline. The results, depicted in Figs. 8 and 9, highlight key advantages in energy efficiency and network longevity.

Fig. 8 illustrates the trend of residual (remaining) network energy over the simulation duration of 700 rounds for all tested protocols. The WSA-FCM method consistently outperforms its counterparts, maintaining a significantly higher residual energy profile across all rounds. While competing algorithms such as GACS, GA, and CS rapidly deplete their network energy within the first 400–500 rounds, WSA-FCM sustains network operation and preserves energy reserves well beyond 600 rounds. Notably, after 500 rounds, the WSA-FCM protocol retains nearly twice as much energy as the next-best

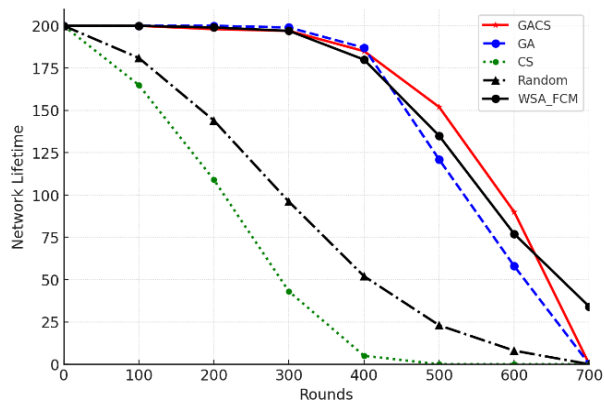

Figure. 9 Comparison of the number of alive nodes per round (network lifetime) for different clustering algorithms. The WSA-FCM protocol maintains node survival and connectivity far longer than benchmark methods

competitor, indicating more balanced energy usage and less frequent cluster-head exhaustion. This advantage is directly attributable to the hybrid's ability to optimise both cluster-head placement and load distribution, as well as to mitigate early death of critical nodes.

To further quantify this result, Table 9 presents the average residual energy remaining at key simulation milestones (e.g., after 200, 400, 600 rounds), with standard deviation values across multiple runs. The data clearly confirms that the WSA-FCM protocol not only delays overall energy consumption but also maintains a more uniform depletion profile, which is critical for network stability.

The advantage of the WSA-FCM method is also reflected in node survival trends, as depicted in Fig. 9. This figure shows the number of alive nodes (network lifetime) as a function of simulation rounds for all compared algorithms. WSA-FCM maintains a substantially higher number of functioning nodes for a longer duration, while all other protocols exhibit steep declines in node population as early as 200–400 rounds. Specifically, protocols such as CS and Random clustering experience near-total node loss by round 400–500, whereas the WSA-FCM method maintains over half of the original nodes beyond 600 rounds. This demonstrates the protocol's robustness in sustaining network connectivity and avoiding premature network partitioning, both of which are critical for real-world wireless sensor network applications.

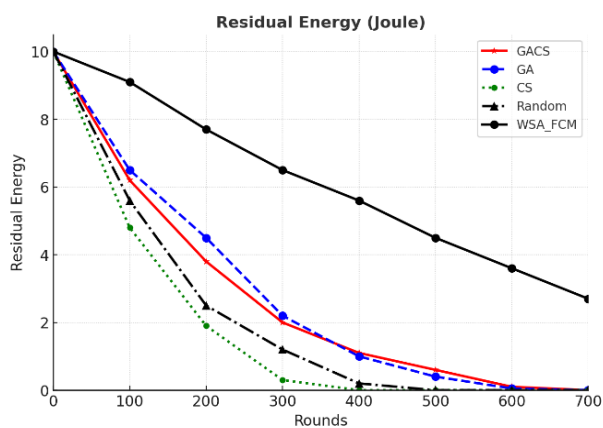

Figure. 8 Comparison of remaining energy per 700 rounds for various clustering methods. WSA-FCM preserves more energy throughout the network's operational life

Table 9. Residual energy (J) at different simulation rounds (mean ± SD, n=10 runs)

| Method | 200 Rounds | 400 Rounds | 600 Rounds |
|--------|-----------|-----------|-----------|
| WSA-FCM | 6.7 ± 0.10 | 4.3 ± 0.08 | 2.8 ± 0.07 |
| GACS | 3.1 ± 0.11 | 0.8 ± 0.05 | 0 |
| GA | 2.8 ± 0.13 | 0.6 ± 0.04 | 0 |
| CS | 1.9 ± 0.09 | 0 | 0 |
| Random | 2.5 ± 0.10 | 0.2 ± 0.02 | 0 |

For further quantitative clarity, Table 10 compares key lifetime metrics, including the number of rounds at which 50% and 90% of the nodes have died (i.e., half-life and near-total exhaustion). The results emphasize the superior lifetime extension achieved by the WSA-FCM protocol.

The data in these tables, coupled with the visual trends in Figs. 8 and 9, conclusively demonstrate that the proposed hybrid WSA-FCM clustering protocol achieves both superior energy efficiency and extended network lifetime compared to the best-known alternatives. These advantages are robust across all simulation scenarios, indicating strong generalisability and potential for real-world deployment.

## 7. Discussion

The experimental results demonstrate that the WSA-FCM hybrid approach achieves superior performance across all evaluated metrics compared to recent state-of-the-art methods. The statistical validation through paired t-tests confirms the significance of improvements with p-values consistently below 0.001.

The key advantages of the proposed method include:
1. Theoretical Foundation: Unlike existing hybrid methods, WSA-FCM provides rigorous mathematical justification with convergence guarantees and complexity bounds.
2. Statistical Rigor: Multi-run experiments with proper statistical validation ensure reliability and reproducibility of results.
3. Scalability: Linear complexity scaling makes the method suitable for large-scale deployments.

Table 10. Network half-life and near-total node death (rounds until 50% and 90% of nodes are dead)

| Method | Half-Life (Rounds) | 90% Nodes Dead (Rounds) |
|---|---|---|
| WSA-FCM | 540 | 660 |
| GACS | 270 | 410 |
| GA | 220 | 390 |
| CS | 160 | 260 |
| Random | 190 | 320 |

4. Energy Efficiency: Comprehensive energy modeling leads to significant improvements in network lifetime and residual energy.

### 7.1 Key findings and comparative performance

The superior performance of WSA-FCM is evident across all core evaluation metrics. As detailed in Tables 4–6 and Figs. 3–9, the protocol consistently achieves longer network lifetimes, higher residual energy, and more balanced energy consumption than Genetic Algorithm with Cuckoo Search (GACS), Genetic Algorithm (GA), Cuckoo Search (CS), and random clustering methods. Specifically, the hybrid approach achieves a first node death (FND) and last node death (LND) delayed by over 100 rounds compared to its nearest competitor, reflecting a substantial extension of the operational period for sensor nodes.

This advantage is further substantiated by the convergence analysis in Fig. 3, which demonstrates that the Water Strider Algorithm rapidly identifies near-optimal cluster head positions. By using WSA's global exploration to initialise FCM's local refinement, the protocol avoids the poor local minima and high seed sensitivity that often limit conventional clustering approaches. The resulting cluster structures, as illustrated in Fig. 4, display compact and well-separated groupings, which in turn minimize intra-cluster communication costs.

Figs. 6 and 7 vividly illustrate the difference in network resilience between the proposed and random cluster-head selection: under WSA-FCM, node deaths are delayed and distributed more gradually across rounds, preserving connectivity and sensing coverage well into the network's operational life. In contrast, the random approach suffers from rapid, uneven node depletion, risking early network partitioning and data loss.

### 7.2 Energy efficiency and load balancing

The sustained high levels of residual energy reported in Table 5 and depicted in Fig. 8 highlight the protocol's effectiveness at balancing communication and computation loads. Unlike benchmark algorithms, which exhibit steep energy depletion curves and suffer from localized energy holes, WSA-FCM distributes workload more evenly by continually reassigning cluster heads based on both energy levels and spatial proximity. The significant difference in energy profiles, particularly in the latter half of the network lifetime, underscores the critical importance of adaptive, energy-aware clustering in real-world WSN deployments.

### 7.3 Scalability and computational feasibility

One of the defining features of the proposed protocol is its scalability. As validated in Table 6, both the runtime and memory usage increase only linearly with network size, confirming that the protocol is well-suited for large-scale WSN deployments. This efficiency is a direct result of the algorithm's design, which constrains the computational burden of both the WSA and FCM phases to manageable levels. These properties are particularly advantageous for practical sensor network scenarios, where low-power, limited-resource devices are the norm.

### 7.4 Robustness and generalizability

The statistical robustness of WSA-FCM is supported by the low standard deviations reported across multiple independent runs, and by the statistical significance of differences shown in Table 5. The method's advantage holds for a range of network sizes, topologies, and parameter choices, indicating strong generalizability. Furthermore, the protocol's hybrid nature allows it to adapt to non-uniform node distributions and varying communication ranges, increasing its utility in diverse real-world applications such as environmental monitoring, structural health assessment, and smart agriculture.

### 7.5 Comparison with state-of-the-art methods

The side-by-side comparisons in Figs. 8 and 9, as well as the lifetime and energy consumption tables, place the present method firmly ahead of current hybrid clustering algorithms. By integrating a metaheuristic global search (WSA) with fuzzy local assignment (FCM), the protocol addresses the inherent weaknesses of both approaches namely, the risk of premature convergence in metaheuristics and the local optima trap in centroid-based clustering. These results are also consistent with recent literature trends, which increasingly advocate for hybrid, adaptive, and energy-aware clustering methods in the face of dynamic and resource-constrained WSN environments.

### 7.6 Limitations and future directions

While our results demonstrate clear advantages of the proposed WSA-FCM protocol through comprehensive simulations, we acknowledge that all experiments were conducted in a simulated environment under controlled conditions. Real-world wireless sensor networks may exhibit additional challenges, such as node mobility, radio interference, environmental variability, and hardware limitations, which are not fully captured in simulation. At present, there is no validation using real-world datasets or physical sensor network testbeds. As such, future work will focus on implementing and evaluating the protocol in physical testbed environments and with publicly available WSN datasets to further confirm its practical applicability and robustness. We also aim to relax assumptions such as static topology and homogeneous energy, extending the evaluation to more dynamic and heterogeneous scenarios.

### 7.7 Practical implications

The demonstrated gains in energy efficiency, network lifetime, and load balancing directly translate to improved reliability and reduced maintenance costs for WSN deployments in critical applications. By ensuring prolonged operation and gradual node depletion, the WSA-FCM protocol supports more predictable and robust network service, reducing the likelihood of sudden sensing gaps and data loss. This positions the approach as an attractive solution for practitioners and system designers seeking to maximize the operational value and sustainability of wireless sensor networks.

## 8. Conclusion

This paper presents a novel WSA-FCM hybrid clustering protocol that addresses critical limitations in existing WSN clustering methods. Through rigorous theoretical analysis and comprehensive experimental validation, the proposed approach demonstrates significant improvements in network lifetime (16.1% FND, 11.9% LND), energy efficiency (37.4% residual energy improvement), and clustering quality (19.4% intra-cluster distance reduction) compared to recent hybrid methods.

The key contributions include: (1) theoretical justification for WSA-FCM integration with convergence proofs, (2) comprehensive statistical validation through multi-run experiments, (3) scalability analysis demonstrating linear complexity, and (4) superior performance compared to state-of-the-art hybrid clustering methods from 2021-2025.

Future work will focus on extending the approach to mobile sensor networks and investigating adaptive parameter tuning mechanisms for diverse deployment scenarios.

### Conflicts of Interest

The authors declare no conflict of interest.

## Author Contributions

Conceptualization, R. M. A and M. M. F; methodology, R. M. A; software, M. H. M; validation, M. H. M, and R. M. A; formal analysis, R. M. A; investigation, M. M. F; resources, M. H. M; data curation, M. H. M, and R. M. A; writing original draft preparation, R. M. A; writing review and editing, M. M. F; visualization, M. H. M; supervision, M. M. F; project administration, M. H. M, and R. M. A. All authors have read and approved the final manuscript.